\documentclass[10pt, final, conference, letterpaper, onecolumn, oneside]{./IEEEtran}

\usepackage{cite}
\usepackage{graphicx}
\usepackage{subfigure}
\usepackage{amsmath}
\usepackage{amssymb}
\usepackage{mathtools}
\usepackage{bm}
\usepackage{epsfig}
\usepackage{times}
\usepackage{color}
\usepackage{url}
\usepackage{array}
\usepackage{multirow}
\usepackage{rotating}
\usepackage{./clrscode_mod}

\newtheorem{pavikc}{\textbf{Corollary}}
\newtheorem{pavikl}{\textbf{Lemma}}
\newtheorem{pavikt}{\textbf{Theorem}}

\newtheorem{pavikp}{\textbf{Proposition}}

\newcommand{\argmax}{\operatornamewithlimits{argmax}}
\newcommand{\pd}[2]{\frac{\partial#1}{\partial#2}}

\begin{document}

\title{Analog Network Coding in General SNR Regime: Performance of a Greedy Scheme}%
\author{\IEEEauthorblockN{Samar Agnihotri, Sidharth Jaggi, and Minghua Chen}\\%
\IEEEauthorblockA{Department of Information Engineering, The Chinese University of Hong Kong, Hong Kong\\}%
Email: samar.agnihotri@gmail.com, \{jaggi, minghua\}@ie.cuhk.edu.hk%
}

\maketitle

\begin{abstract}
The problem of maximum rate achievable with analog network coding for a unicast communication over a layered relay network with directed links is considered. A relay node performing analog network coding scales and forwards the signals received at its input. Recently this problem has been considered under certain assumptions on per node scaling factor and received SNR. Previously, we established a result that allows us to characterize the optimal performance of analog network coding in network scenarios beyond those that can be analyzed using the approaches based on such assumptions.

The key contribution of this work is a scheme to greedily compute a lower bound to the optimal rate achievable with analog network coding in the general layered networks. This scheme allows for exact computation of the optimal achievable rates in a wider class of layered networks than those that can be addressed using existing approaches. For the specific case of Gaussian $N$-relay diamond network, to the best of our knowledge, the proposed scheme provides the first exact characterization of the optimal rate achievable with analog network coding. Further, for general layered networks, our scheme allows us to compute optimal rates within a constant gap from the cut-set upper bound asymptotically in the source power.
\end{abstract}

\section{Introduction}
\label{sec:intro}
Analog network coding (ANC) extends to multihop wireless networks the idea of linear network coding \cite{103liYeungCai}, where an intermediate node sends out a linear combination of its incoming packets. In a wireless network, signals transmitted simultaneously by multiple sources add in the air. Each node receives a \textit{noisy sum} of these signals, \textit{i.e.} a linear combination of the received signals and noise. A communication scheme wherein each relay node merely amplifies and forwards this noisy sum is referred to as analog network coding \cite{107kattiGollakottaKatabi, 110maricGoldsmithMedard}.

The rates achievable with ANC in layered relay networks is analyzed in \cite{110maricGoldsmithMedard, 111liuCai}. In \cite{110maricGoldsmithMedard}, the achievable rate is computed under two assumptions: (A) each relay node scales the received signal to the maximum extent subject to its transmit power constraint, (B) the nodes in all $L$ layers operate in the high-SNR regime, where $\min_{k \in l} P_{R,k} \ge 1/\delta, l = 1, \ldots, L$ for $\delta \ge 0$, where $P_{R,k}$ is the received signal power at the $k^\textrm{th}$ node. It is shown that the rate achieved under these two assumptions approaches network capacity as the source power increases. The authors in \cite{111liuCai} extend this result to the scenarios where the nodes in at most one layer do not satisfy these assumptions and show that achievable rates in such scenarios still approach the network capacity as the source power increases\footnote{However, it is assumed that the noises at the nodes in this particular layer are independent, resulting in the computed ANC rate overestimating the optimal ANC rate in general.}.

However, requiring each relay node to amplify its received signal to the upper bound of its transmit power constraint results in suboptimal end-to-end performance of analog network coding, as we show in \cite{111agnihotriJaggiChen, 112agnihotriJaggiChen}. Further, even in low-SNR regimes amplify-and-forward relaying can be capacity-achieving relay strategy in some scenarios, \cite{107gomadamJafar}.

In this paper we are concerned with analyzing the performance of analog network coding in general layered networks, without the above two assumptions on input signal scaling factors and received SNRs. However, such a characterization of the performance of analog network coding results in a computationally intractable problem in general \cite{111liuCai, 111agnihotriJaggiChen}.

In \cite{112agnihotriJaggiChen}, we establish that a globally optimal set of scaling factors for each node, {\it i.e.} a choice of relaying strategies that optimizes end-to-end throughput over all ANC strategies, can be computed in a layer-by-layer manner. This result allows us to computationally efficiently characterize exactly the optimal ANC rate in a large class of layered networks that cannot be addressed using existing approaches under the assumptions A and B. Further, for general layered relay networks, this result significantly reduces the computational complexity of computing a set of non-trivial achievable rates.

However, even layer-by-layer computation of a network-wide scaling vector that maximizes the end-to-end ANC rate for general layered networks is a computationally hard problem. In this paper, we propose a greedy scheme to bound from below the optimal rate achievable with analog network coding in general layered networks. The proposed scheme allows us to exactly compute the optimal ANC rate in a much wider class of layered networks than those that can be so addressed using existing approaches, including our approach in \cite{112agnihotriJaggiChen}. In particular, for the Gaussian $N$-relay diamond network \cite{111nazarogluOzgurFragouli}, the proposed scheme allows us to exactly compute the optimal rate achievable with analog network coding. To the best of our knowledge, this is the first characterization of the optimal ANC rate for Gaussian diamond network. Further, for general layered networks, our scheme allows for the computation of the optimal rates within a constant gap from the cut-set upper bound asymptotically in the source power.

\textit{Organization:} In Section~\ref{sec:sysModel} we introduce a general wireless layered relay network model and formulate the problem of maximum rate achievable with ANC in such a network. Section~\ref{sec:diamond} addresses the problem of maximum ANC rate achievable in a Gaussian $N$-relay diamond network and shows that a greedy scheme optimally solves this problem. In Section~\ref{sec:genNet} we first generalize the greedy scheme for Gaussian diamond networks to characterize the optimal performance of a specific subnetwork of the general layered network. We then construct a scheme to bound from below the optimal performance of ANC in general layered networks. Section~\ref{sec:exa} illustrates that the proposed scheme leads to the exact computation of the maximum ANC rate in a specific class of layered networks and tight characterization of the optimal rate in the general layered networks asymptotically in the source power. Section~\ref{sec:conclFW} concludes the paper.

\section{System Model}
\label{sec:sysModel}
\begin{figure}[!t]
\centering
\includegraphics[width=3.0in]{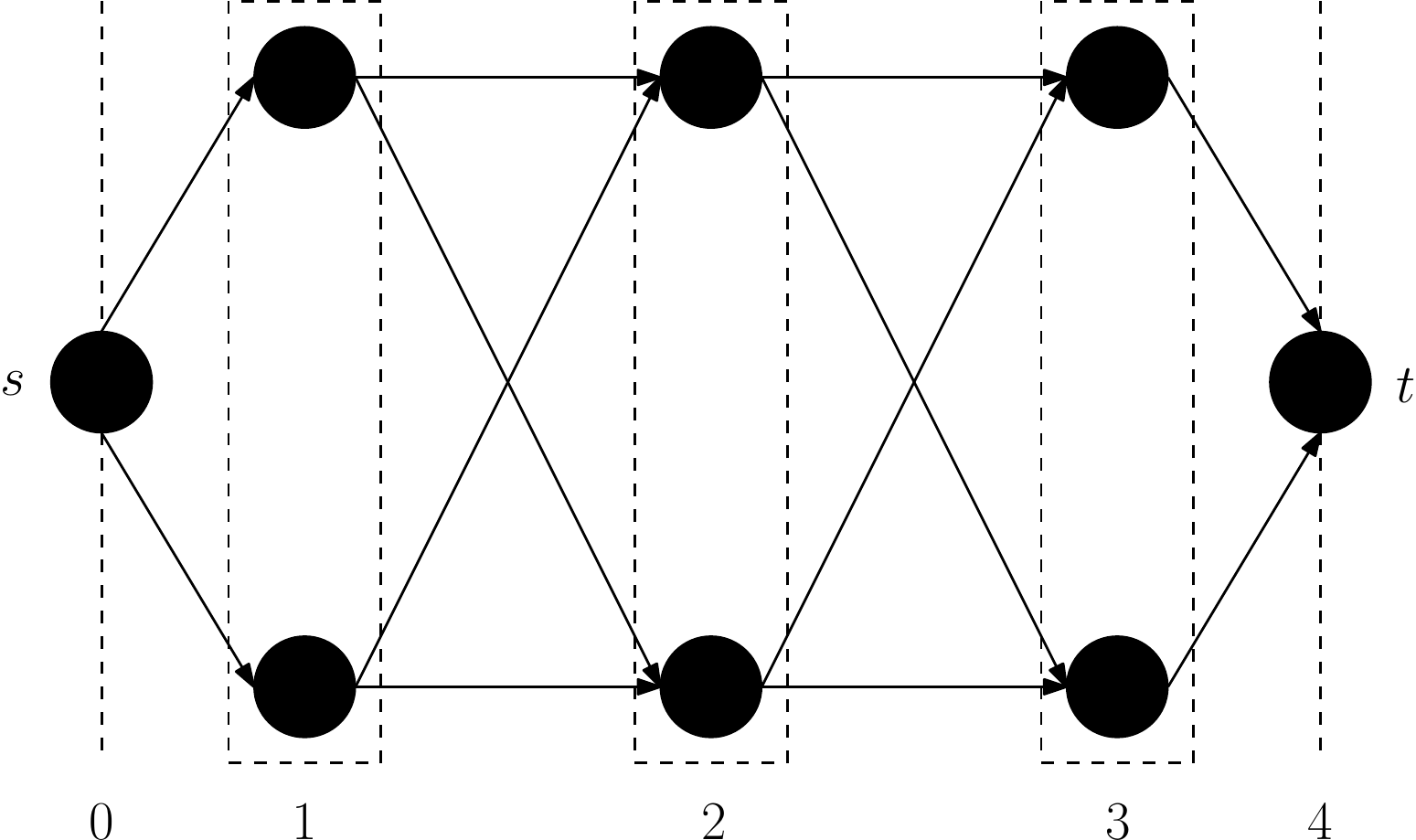}
\caption{Layered network with 3 relay layers between the source `s' and destination `t'. Each layer contains two relay nodes.}
\label{fig:layrdNetExa}
\end{figure}

Consider a $(L+2)$-layer wireless network with directed links. The source $s$ is at the layer `$0$', the destination $t$ is at the layer `$L+1$', and the relay nodes from the set $R$ are arranged in $L$ layers between them. The $l^\textrm{th}$ layer contains $n_l$ relay nodes, $\sum_{l=1}^L n_l = M$. An instance of such a network is given in Figure~\ref{fig:layrdNetExa}. Each node is assumed to have a single antenna and operate in full-duplex mode.

At instant $n$, the channel output at node $i, i \in R \cup \{t\}$, is
\begin{equation}
\label{eqn:channelOut}
y_i[n] = \sum_{j \in {\mathcal N}(i)} h_{ji} x_j[n] + z_i[n], \quad - \infty < n < \infty,
\end{equation}
where $x_j[n]$ is the channel input of the node $j$ in the neighbor set ${\mathcal N}(i)$ of node $i$. In \eqref{eqn:channelOut}, $h_{ji}$ is a real number representing the channel gain along the link from node $j$ to node $i$. It is assumed to be fixed (for example, as in a single realization of a fading process) and known throughout the network. The source symbols $x_s[n], - \infty < n < \infty$, are i.i.d. Gaussian random variables with zero mean and variance $P_s$ that satisfy an average source power constraint, $x_s[n] \sim {\cal N}(0, P_s)$. Further, $\{z_i[n]\}$ is a sequence (in $n$) of i.i.d. Gaussian random variables with $z_i[n] \sim {\cal N}(0, \sigma^2)$. We also assume that $z_i$ are independent of the input signal and of each other. We assume that the $i^{\textrm{th}}$ relay's transmit power is constrained as:
\begin{equation}
\label{eqn:pwrConstraint}
E[x_i^2[n]] \le P_i, \quad - \infty < n < \infty
\end{equation}

In analog network coding each relay node amplifies and forwards the noisy signal sum received at its input. More precisely, a relay node $i$ at instant $n+1$ transmits the scaled version of $y_i[n]$, its input at time instant $n$, as follows
\begin{equation}
\label{eqn:AFdef}
x_i[n+1] = \beta_i y_i[n], \quad 0 \le \beta_i^2 \le \beta_{i,max}^2 = P_i/P_{R,i},
\end{equation}
where $P_{R,i}$ is the received power at the node $i$ and choice of the scaling factor $\beta_i$ satisfies the power constraint \eqref{eqn:pwrConstraint}.

One important characteristic of layered networks with unidirectional links is that all paths from the source to destination have same number of hops. Also, each path from the $i^\textrm{th}, i \in R$, relay node to the destination has the same length. Therefore, in a layered network with $L$ layers, all copies of a source signal traveling along different paths arrive at the destination with time delay $L$ and all copies of a noise symbol introduced at a node in $l^\textrm{th}$ layer arrive at the destination with time delay $L-i+1$. Therefore, the outputs of the source-destination channel are free of intersymbol interference. This simplifies the relation between input and output of the source-destination channel and allows us to omit the time-index while denoting the input and output signals.

Using \eqref{eqn:channelOut} and \eqref{eqn:AFdef}, the input-output channel between the source and destination can be written as
\begin{equation}
\label{eqn:sdchnl}
y_t =  \bigg[\sum_{(i_1, \ldots, i_{L}) \in K_{L}} \hspace{-0.25in} h_{si_1}\beta_{i_1}h_{i_1 i_2} \ldots \beta_{i_{L}}h_{i_{L} t}\bigg] x_s +  \sum_{l=1}^{L} \sum_{j=1}^{n_l}\bigg[\sum_{(i_l, \ldots, i_{L}) \in K_{lj,L}} \hspace{-0.25in} \beta_{lj} h_{lj, i_l} \ldots \beta_{L i_{L}}h_{L i_{L}, t}\bigg] z_{lj} + z_t, \nonumber
\end{equation}
where $K_L$, is the set of $L$-tuples of node indices corresponding to all paths from the source to the destination with path delay $L$. Similarly, $K_{lj,L-l+1}$, is the set of $L-l+1$-tuples of node indices corresponding to all paths from the $j^{\textrm{th}}$ relay of $l^\textrm{th}$ layer to the destination with path delay $L-l+1$.

We introduce \textit{modified} channel gains as follows. For all the paths between the source $s$ and the destination $t$:
\begin{equation}
\label{eqn:modChnlParams}
h_s = \sum_{(i_1, \ldots, i_{L}) \in K_{L}} h_{si_1}\beta_{i_1}h_{i_1 i_2} \ldots \beta_{i_{L}}h_{i_{L} t}
\end{equation}
For all the paths between the $j^{\textrm{th}}$ relay of $l^\textrm{th}$ layer to the destination $t$ with path delay $L-l+1$:
\begin{equation}
\label{eqn:modChnlParams2}
h_{lj} = \sum_{(i_1, \ldots, i_{L-l+1}) \in K_{lj,L-l+1}} \beta_{lj} h_{lj, i_1} \ldots \beta_{L i_{L}}h_{L i_{L}, t}
\end{equation}

In terms of these modified channel gains\footnote{Modified channel gains for even a possibly exponential number of paths as in \eqref{eqn:modChnlParams} and \eqref{eqn:modChnlParams2} can be efficiently computed using line-graphs \cite{103koetterMedard}. Further, the number of such modified channel gains scales polynomially in the size of the graph being considered.}, the source-destination channel in \eqref{eqn:sdchnl} can be written as:
\begin{equation}
\label{eqn:chnlmod}
y_t = h_s x_s + \sum_{l=1}^{L} \sum_{j=1}^{n_l} h_{lj} z_{lj} + z_t 
\end{equation}

\textit{Problem Formulation:} For a given network-wide scaling vector $\bm{\beta}=(\beta_{li})_{1 \le l \le L, 1 \le i \le n_l}$, the achievable rate for the channel in \eqref{eqn:chnlmod} with i.i.d. Gaussian input is (\hspace{-0.001cm}\cite{110maricGoldsmithMedard, 111liuCai, 111agnihotriJaggiChen}):
\begin{equation}
\label{eqn:infoRateFin}
I(P_s, \bm{\beta}) = (1/2) \log\big(1 + SNR_t\big),
\end{equation}
where $SNR_t$, the signal-to-noise ratio at the destination $t$ is:
\begin{equation}
\label{eqn:snr}
SNR_t = \frac{P_s}{\sigma^2}\frac{h_s^2}{1 + \sum_{l=1}^{L} \sum_{j=1}^{n_l} h_{lj}^2}
\end{equation}

The maximum information-rate $I_{ANC}(P_s)$ achievable in a given layered network with i.i.d. Gaussian input is defined as the maximum of $I(P_s, \bm{\beta})$ over all feasible $\bm{\beta}$, subject to per relay transmit power constraint \eqref{eqn:AFdef}. In other words:
\begin{equation}
\label{eqn:maxAFrate}
I_{ANC}(P_s) \stackrel{def}{=} \max_{\bm{\beta}:0 \le \beta_{li}^2 \le \beta_{li, max}^2} I(P_s, \bm{\beta})
\end{equation}
It should be noted that $\beta_{li, max}$ (the maximum value of the scaling factor for $i^\textrm{th}$ node in the $l^\textrm{th}$ layer) depends on the scaling factors for the nodes in the previous $l-1$ layers.

Given the monotonicity of the $\log(\cdot)$ function, we have
\begin{equation}
\label{eqn:eqProb}
\bm{\beta}_{opt} = \argmax_{\bm{\beta}:0 \le \beta_{li}^2 \le \beta_{li, max}^2}  I(P_s, \bm{\beta}) = \argmax_{\bm{\beta}:0 \le \beta_{li}^2 \le \beta_{li, max}^2}  SNR_t
\end{equation}
Therefore in the rest of the paper, we concern ourselves mostly with maximizing the received SNRs.

In \cite{112agnihotriJaggiChen}, we discussed the computational complexity of exactly solving the problem \eqref{eqn:maxAFrate} or equivalently the problem \eqref{eqn:eqProb}. Further, we also introduced a key result \cite[Lemma 2]{112agnihotriJaggiChen} that reduces the computational complexity of the problem of computing $\bm{\beta}_{opt}$ by computing it layer-by-layer as a solution of a cascade of subproblems. This result allowed us to characterize the optimal end-to-end rate achievable with analog network coding in communication scenarios that cannot be so addressed using previous approaches. However, each of these subproblems itself is computationally hard for general network scenarios as it involves maximizing the ratio of \textit{posynomials} \cite{107boydkimVandenberghe, 105chiang}, which is known to be computationally intractable in general \cite{105chiang}. Therefore, in this paper, we introduce a greedy scheme that optimally solves these subproblems and consequently the problem \eqref{eqn:eqProb} for a large class of layered networks that cannot be addressed with current schemes. For general layered networks, the proposed scheme allows us to tightly bound from below the optimal ANC performance. However before discussing this scheme, we motivate it by computing the maximum rate of information transfer achievable with analog network coding over the diamond network with $N$ relay nodes.

\section{Diamond Network: The optimal rate achievable with analog network coding}
\label{sec:diamond}
\begin{figure}[!t]
\centering
\includegraphics[width=3.0in]{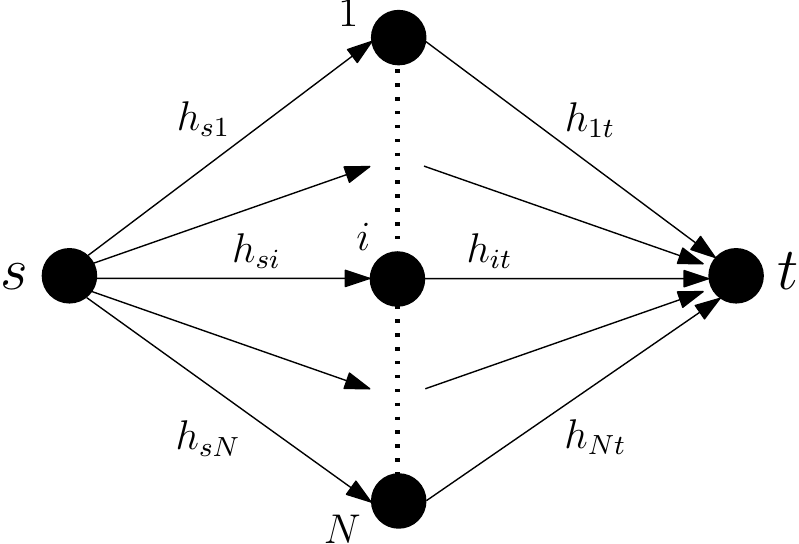}
\caption{A diamond network with $N$ relay nodes.}
\label{fig:diamond}
\end{figure}

Consider the diamond network of Figure~\ref{fig:diamond}. We can consider diamond network as a layered network with only one layer of relay nodes. Then using \eqref{eqn:modChnlParams}, \eqref{eqn:modChnlParams2}, and \eqref{eqn:snr}, we compute the SNR at the destination $t$ for any scaling vector $\bm{\beta} = (\beta_1, \ldots, \beta_N)$ as
\begin{equation}
\label{eqn:diamondSNR}
SNR_t = \frac{P_s}{\sigma^2} \frac{(\sum_{i=1}^N h_{si} \beta_i h_{it})^2}{1+\sum_{i=1}^N \beta_i^2 h_{it}^2}
\end{equation}

Using \eqref{eqn:maxAFrate}, the problem of computing the maximum ANC rate for this network thus can be formulated as
\begin{equation}
\label{eqn:diamondProb}
\max_{0 \le \bm{\beta}^2 \le \bm{\beta}_{max}^2} SNR_t,
\end{equation}
where $\bm{\beta}_{max} = (\beta_{1,max} \ldots, \beta_{N,max})$ with $\beta_{i,max}^2 = P_i/(h_{si}^2 P_s + \sigma^2), i \in {\mathcal N}, {\mathcal N} = \{1, \ldots, N\}$.

Equating the first-order partial derivatives of the objective function with respect to $\beta_i, i \in \mathcal{N}$, to zero, we get the following $N+1$ conditions for local extrema:
\begin{align}
&\sum_{i \in \mathcal{N}} h_{si} \beta_i h_{it} = 0 \label{eqn:b1bN} \\
&\beta_i = \frac{h_{si}/h_{it}}{\sum_{j \in \mathcal{N}\setminus\{i\}} h_{sj} \beta_j h_{jt}}\bigg(1+\sum_{j \in \mathcal{N}\setminus\{i\}} \beta_j^2 h_{it}^2\bigg) \label{eqn:bi_ito_rest}
\end{align}

Let $SNR_{\beta_i \beta_j} = \frac{\partial^2 SNR_t}{\partial \beta_i \partial \beta_j}$ denote the second-order partial derivatives of $SNR_t$ with respect to $\beta_i$ and $\beta_j$, $i, j \in \mathcal{N}$ and $H(\bm{\beta})$ denote the determinant of $N \times N$ Hessian matrix.

First, consider the set of stationary points $S_{\bm{\beta}} = \{\bm{\beta}: \bm{\beta} \mbox{ satisfies } \eqref{eqn:b1bN}\}$. For all points in $S_{\bm{\beta}}$ we can prove that
\begin{align*}
SNR_{\beta_1 \beta_1} &> 0\\
H(\bm{\beta}) &= 0
\end{align*}
Therefore, the second partial derivative test to determine if the points in $S_{\bm{\beta}}$ are local minimum, maximum, or saddle points fails. However, we can establish that for every $\bm{\beta} \in S_{\bm{\beta}}$, the following set of conditions holds
\begin{align}
\pd{SNR_t}{\beta_i}\bigg|_{\bm{\beta}+\bm{\delta}} &< 0, \mbox{ if } \sum_{i \in \mathcal{N}} h_{si} h_{it} \delta_i < 0, \label{eqn:ineq1}\\
\pd{SNR_t}{\beta_i}\bigg|_{\bm{\beta}+\bm{\delta}} &> 0, \mbox{ if } \sum_{i \in \mathcal{N}} h_{si} h_{it} \delta_i > 0, \label{eqn:ineq2}\\
H(\bm{\beta}) &> 0, \mbox{ if } \sum_{i \in \mathcal{N}} h_{si} h_{it} \delta_i < 0, \label{eqn:ineq3}\\
H(\bm{\beta}) &> 0, \mbox{ if } \sum_{i \in \mathcal{N}} h_{si} h_{it} \delta_i > 0, \label{eqn:ineq4}
\end{align}
for all $\bm{\delta}=(\delta_1, \ldots, \delta_N) \rightarrow \bm{0}$. In other words, \eqref{eqn:ineq1} and \eqref{eqn:ineq2} imply that the slope of the function changes sign at $\sum_{i \in \mathcal{N}} h_{si} h_{it} \delta_i = 0$, and \eqref{eqn:ineq3} and \eqref{eqn:ineq4} imply that the convexity of the function, however, does not change at $\sum_{i \in \mathcal{N}} h_{si} h_{it} \delta_i = 0$. Therefore, together these imply that \eqref{eqn:b1bN} leads to a local minimum of the objective function.

Next, consider the set of points defined by \eqref{eqn:bi_ito_rest}. For all such points we can prove that
\begin{align*}
SNR_{\beta_1 \beta_1} &< 0\\
H(\bm{\beta}) &> 0
\end{align*}
Therefore, from the second partial derivative test the objective function attains its local maximum at the set of points characterized by \eqref{eqn:bi_ito_rest} above. However, no real solution of the simultaneous system of equations in \eqref{eqn:bi_ito_rest} exists. In other words, no solution of \eqref{eqn:diamondProb} exists where all relay nodes transmit strictly below their respective transmit power constraints. This is illustrated by the following example.

{\textit{\textbf{Example 1} (Three node Diamond Network):}} Consider the Gaussian diamond network of Figure~\ref{fig:diamond} with three relay nodes. For this network, \eqref{eqn:bi_ito_rest} results in:
\begin{align}
\beta_1 &= \frac{h_{s1}/h_{1t}}{h_{s2} \beta_2 h_{2t}+h_{s3} \beta_3 h_{3t}}\big(1+ \beta_2^2 h_{2t}^2 + \beta_3^2 h_{3t}^2\big) \label{eqn:beta1diamond3} \\
\beta_2 &= \frac{h_{s2}/h_{2t}}{h_{s1} \beta_1 h_{1t}+h_{s3} \beta_3 h_{3t}}\big(1+ \beta_1^2 h_{1t}^2 + \beta_3^2 h_{3t}^2\big) \label{eqn:beta2diamond3} \\
\beta_3 &= \frac{h_{s3}/h_{3t}}{h_{s1} \beta_1 h_{1t}+h_{s2} \beta_2 h_{2t}}\big(1+ \beta_1^2 h_{1t}^2 + \beta_2^2 h_{2t}^2\big) \label{eqn:beta3diamond3}
\end{align}

Substituting \eqref{eqn:beta2diamond3} in \eqref{eqn:beta3diamond3}, after a little algebraic manipulation we get:
\begin{equation*}
\label{eqn:beta3condition}
\{\beta_3 h_{3t} h_{s1} \beta_1 h_{1t} - h_{s3}(1+\beta_1^2 h_{1t}^2)\}\{(h_{s1} \beta_1 h_{1t}+h_{s3} \beta_3 h_{3t})^2 + h_{s2}^2(1 + \beta_1^2 h_{1t}^2 + \beta_3^2 h_{3t}^2)\}=0
\end{equation*}
Solving this gives three solutions for $\beta_3$, namely:
\begin{align}
\beta_{3,1} &= \frac{h_{03}/h_{3t}}{h_{s1} \beta_1 h_{1t}} (1+\beta_1^2 h_{1t}^2) \label{eqn:beta3sol1} \\
\beta_{3,2} &= \frac{- h_{s3} h_{s1} \beta_1 h_{1t} + i h_{s2} \sqrt{(h_{s2}^2 + h_{s3}^2)(1+\beta_1^2 h_{1t}^2) + (h_{s1} \beta_1 h_{1t})^2}}{h_{3t}(h_{s2}^2 + h_{s3}^2)}, i = \sqrt{-1} \label{eqn:beta3sol2} \\
\beta_{3,2} &= \frac{- h_{s3} h_{s1} \beta_1 h_{1t} - i h_{s2} \sqrt{(h_{s2}^2 + h_{s3}^2)(1+\beta_1^2 h_{1t}^2) + (h_{s1} \beta_1 h_{1t})^2}}{h_{3t}(h_{s2}^2 + h_{s3}^2)}, i = \sqrt{-1} \label{eqn:beta3sol3}
\end{align}
Substituting each of \eqref{eqn:beta3sol1}, \eqref{eqn:beta3sol2}, and \eqref{eqn:beta3sol3} in \eqref{eqn:beta2diamond3} results in three corresponding solutions for $\beta_2$, namely:
\begin{align*}
\beta_{2,1} &= \frac{h_{02}/h_{2t}}{h_{s1} \beta_1 h_{1t}} (1+\beta_1^2 h_{1t}^2) \\
\beta_{2,2} &= \frac{- h_{s2} h_{s1} \beta_1 h_{1t} + i h_{s3} \sqrt{(h_{s2}^2 + h_{s3}^2)(1+\beta_1^2 h_{1t}^2) + (h_{s1} \beta_1 h_{1t})^2}}{h_{2t}(h_{s2}^2 + h_{s3}^2)}, i = \sqrt{-1} \\
\beta_{2,3} &= \frac{- h_{s2} h_{s1} \beta_1 h_{1t} - i h_{s3} \sqrt{(h_{s2}^2 + h_{s3}^2)(1+\beta_1^2 h_{1t}^2) + (h_{s1} \beta_1 h_{1t})^2}}{h_{2t}(h_{s2}^2 + h_{s3}^2)}, i = \sqrt{-1}
\end{align*}
Therefore, we have three possible solutions for optimal $(\beta_2, \beta_3)$, \textit{i.e.} $(\beta_{2,1}, \beta_{3,1})$, $(\beta_{2,2}, \beta_{3,2})$, and $(\beta_{2,3}, \beta_{3,3})$.

Substituting $(\beta_{2,1}, \beta_{3,1})$ in \eqref{eqn:beta1diamond3} results in
\begin{equation*}
\beta_1^2 = -\frac{h_{s2}^2 + h_{s3}^2}{h_{1t}^2 (h_{s1}^2 + h_{s2}^2 + h_{s3}^2)},
\end{equation*}
which leads to complex valued solutions for optimal $\beta_1$.

Similarly, substituting $(\beta_{2,2}, \beta_{3,2})$, and $(\beta_{2,3}, \beta_{3,3})$ in \eqref{eqn:beta1diamond3} results in
\begin{equation*}
\beta_1^2 = -\frac{1 + h_{s2}^2 + h_{s3}^2}{h_{1t}^2 (h_{s2}^2 + h_{s3}^2)}
\end{equation*}
which also leads to complex valued solutions for optimal $\beta_1$.

This allows us to conclude that no real solution of the system of simultaneous equations in \eqref{eqn:beta1diamond3}-\eqref{eqn:beta3diamond3} exists.  {\hspace*{\fill}~\IEEEQEDclosed\par}

The above discussion implies that all points satisfying \eqref{eqn:b1bN} lead to the global minimum of the objective function in \eqref{eqn:diamondProb} and the global maximum of the objective function occurs at one of the $N$ hyperplanes (of dimension $N-1$) defined by $\beta_k = \beta_{k,max}, k \in \mathcal{N}$. Next we identify this hyperplane and characterize the corresponding optimal solution.

Consider the system of simultaneous equations in \eqref{eqn:bi_ito_rest} on the $(N-1)$-dimensional hyperplane defined by $\beta_k = \beta_{k,max}$.
\begin{equation}
\label{eqn:bi_ito_rest_on_kth_plane}
\beta_i = \frac{h_{si}}{h_{it}} \frac{1+\beta_{k,max}^2 h_{kt}^2+\sum_{j \in \mathcal{N}\setminus\{i,k\}} \beta_j^2 h_{jt}^2}{h_{sk} \beta_{k,max} h_{kt} + \sum_{j \in \mathcal{N}\setminus\{i,k\}} h_{sj} \beta_j h_{jt}}
\end{equation}
Note that the solution of the above system of equations is the set of scaling-factors for the nodes in the set of relay nodes $\mathcal{N} \setminus \{k\}$ that maximizes $SNR_t$ on hyperplane $\beta_k = \beta_{k,max}$. Solving the system of equations in \eqref{eqn:bi_ito_rest_on_kth_plane} results in the following set of optimal solutions for $\beta_i$ on hyperplane $\beta_k = \beta_{k,max}$:
\begin{equation}
\label{eqn:optBeta_on_kth_plane}
\beta_{i}^k = \frac{h_{si}}{h_{it}} \frac{1+\beta_{k,max}^2 h_{kt}^2}{h_{sk} \beta_{k,max} h_{kt}}, i \in \mathcal{N}\setminus \{k\}
\end{equation}

However, the optimal scaling factors in \eqref{eqn:optBeta_on_kth_plane} for $N-1$ nodes are computed without considering the upper bound $\beta_{i,max}$ on each $\beta_i, i \in \mathcal{N}\setminus\{k\}$. Therefore, taking into consideration the upper bound on the scaling factor for each node, the modified solution is computed as per the following lemma.

\begin{pavikl}
\label{lemma:optBeta}
The optimal scaling vector $\bm{\beta}_{opt}^k=(\beta_{1,opt}^k, \ldots, \beta_{N,opt}^k)$ for $N$ nodes on $\beta_k = \beta_{k,max}$ hyperplane such that each scaling factor satisfies the corresponding upper bound on its maximum value is given as
\begin{equation*}
\beta_{i,opt}^k = \begin{dcases}
                   \beta_{i,max}, i \in S^k \\
                     \frac{h_{si}}{h_{it}} \frac{1+\sum_{j \in S^k} \beta_{j,max}^2 h_{jt}^2}{\sum_{j \in S^k} h_{sj} \beta_{j,max} h_{jt}}, i \not\in S^k,
                  \end{dcases}
\end{equation*}
where $S^k$ is the set of nodes such that on hyperplane $\beta_k = \beta_{k,max}$, the optimal value of the scaling factor of a node is saturated to its corresponding upper bound, $S^k = \{k\} \cup \{i: \beta_{i}^k \ge \beta_{i,max}, i \in \mathcal{N}\setminus\{k\}\}$.
\end{pavikl}
\begin{IEEEproof}
Following the argument similar to the one used to prove the global extrema properties of \eqref{eqn:b1bN} and \eqref{eqn:bi_ito_rest}, we can prove that on the $\beta_k = \beta_{k,max}$ hyperplane, the $SNR_t$ achieves its global minimum at a hyperplane defined by
\begin{equation*}
\sum_{i \in \mathcal{N}\setminus\{k\}} h_{si} \beta_i h_{it} = 0
\end{equation*}
and its global maximum at the points defined by $\beta_i^k$ given in \eqref{eqn:optBeta_on_kth_plane}.

Let $M^k$ denotes the set of nodes for which $\beta_i^k$ computed in \eqref{eqn:optBeta_on_kth_plane} is greater than or equal to the corresponding upper bound $\beta_{i,max}$ on the maximum value of the scaling factor, \textit{i.e.} $M^k = \{i: \beta_{i}^k \ge \beta_{i,max}, i \in \mathcal{N}\setminus\{k\}\}$. For all such $\beta_i^k, i \in M^k$, after proving that $\pd{SNR_t}{\beta_i}\big|_{\beta_{i,max}} \ge 0$, we set $\beta_i^k = \beta_{i,max}$ and update $S^k$, the set of nodes such that on hyperplane $\beta_k = \beta_{k,max}$, the optimal value of each node is saturated to its corresponding upper bound; as follows: $S^k = S^k \cup M^k$. As $\beta_i^k$ computed in \eqref{eqn:optBeta_on_kth_plane} for a node $i \not\in S^k$ may no longer be optimal after the above re-assignment of $\beta_i^k, i \in M^k$, we need to solve the following simultaneous system of $N - |S^k| = N-|M^k|-1$ equations with $i \in \mathcal{N} \setminus S^k$:
\begin{equation}
\label{eqn:newSysOeqns}
\beta_i = \frac{h_{si}}{h_{it}} \frac{1+\sum\limits_{j \in S^k} \beta_{j,max}^2 h_{jt}^2+\sum\limits_{j \not\in S^k \cup \{i\}} \beta_{j}^2 h_{jt}^2}{\sum\limits_{j \in S^k} h_{sj} \beta_{j,max} h_{jt} + \sum\limits_{j \not\in S^k \cup \{i\}} h_{sj} \beta_{j} h_{jt}},
\end{equation}
Solving this system of equations results in
\begin{equation}
\label{eqn:recomputedBeta}
\beta_{i,opt}^k = \frac{h_{si}}{h_{it}} \frac{1+\sum_{j \in S^k} \beta_{j,max}^2 h_{jt}^2}{\sum_{j \in S^k} h_{sj} \beta_{j,max} h_{jt}}, i \not\in S^k
\end{equation}
Some of the recomputed scaling factors $\beta_{i,opt}^k, i \not\in S^k$ may violate the corresponding upper bound on their maximum value. All such nodes are added to set $S^k$, thus updating it. Then, the system of equations in \eqref{eqn:newSysOeqns} is solved again for this updated set $S^k$. This iterative process continues until none of the recomputed $\beta_i$ in \eqref{eqn:recomputedBeta} violates its corresponding upper bound. This iterative process is presented formally in terms of an algorithm: \textit{Algorithm 1}, given on the top of the next page.

Note that \textit{Algorithm 1} always halts with either $S^k = \mathcal{N} \setminus \{k\}$ or $S^k \subset \mathcal{N} \setminus \{k\}$ and $\beta_i^k < \beta_{i,max}, i \in \mathcal{N} \setminus S^k$.
\end{IEEEproof}

Using Lemma~\ref{lemma:optBeta}, for each of $N$ hyperplanes, defined as $\beta_k = \beta_{k,max}$, $k \in \mathcal{N}$, we can compute $\bm{\beta}_{opt}^k$, the set of scaling factors for all nodes at which $SNR_t$ attains its maximum on $\beta_k = \beta_{k,max}$ hyperplane. Then the hyperplane at which $SNR_t$ attains its global maximum is identified as follows:

\begin{pavikp}
\label{prop:optHyperplane}
The hyperplane at which $SNR_t$ attains its global maximum is defined as
\begin{equation*}
k^\star = \argmax_{k \in \mathcal{N}} SNR_t(\bm{\beta}_{opt}^k)
\end{equation*}
\end{pavikp}

Combining Proposition~\ref{prop:optHyperplane} and Lemma~\ref{lemma:optBeta}, we can characterize the scaling vector $\bm{\beta}_{opt}$ that solves the problem \eqref{eqn:diamondProb} as follows.

\newpage
\hspace{-1.0em}\hrulefill

\hspace{-1.0em}{\textbf{Algorithm 1}}

\vspace{-0.3cm}\hspace{-1.0em}\hrulefill
\begin{codebox}
\li Initialization:
\zi $S^k = \{k\}$, the set of nodes whose scaling factors are saturated to their respective upper-bounds
\zi on hyperplane $\beta_k = \beta_{k,max}$.
\zi $U^k = \mathcal{N}\setminus\{k\}$ , the set of nodes whose scaling factors are not saturated to their respective upper-bounds
\zi on hyperplane $\beta_k = \beta_{k,max}$.
\li Compute $\beta_{i}^k = {\displaystyle \frac{h_{si}}{h_{it}} \frac{1+\sum_{j \in S^k} \beta_{j,max}^2 h_{jt}^2}{\sum_{j \in S^k} h_{sj} \beta_{j,max} h_{jt}}}, i \in U^k$.
\li \While ($\exists \, \beta_{i}^k \ge \beta_{i,max}, i \in U^k$)
\li \hspace{-0.45in}  Compute $M^k = \{i: \beta_{i}^k \ge \beta_{i,max}, i \in U^k\}$.
\li \hspace{-0.45in}  $\beta_{i}^k = \beta_{i,max}, i \in M^k$.
\li \hspace{-0.45in}  $S^k = S^k \cup M^k$.
\li \hspace{-0.45in}  $U^k = U^k \setminus M^k$.
\li \hspace{-0.45in}  Compute $\beta_{i}^k = {\displaystyle \frac{h_{si}}{h_{it}} \frac{1+\sum_{j \in S^k} \beta_{j,max}^2 h_{jt}^2}{\sum_{j \in S^k} h_{sj} \beta_{j,max} h_{jt}}}, i \in U^k$.
    \End
\end{codebox}
\vspace{-0.2cm}\hrulefill

\begin{pavikt}
\label{thrm:optBeta4diamond}
A network-wide scaling vector $\bm{\beta}_{opt} = (\beta_{1}^{opt}, \ldots, \beta_{N}^{opt})$ that maximizes the $SNR_t$ for a diamond network with the relay nodes performing ANC is given as
\begin{equation*}
\beta_{i}^{opt} = \begin{dcases}
                    \beta_{i,max}, i = k^\star, k^\star = \argmax\limits_{j \in \mathcal{N}}SNR_t(\bm{\beta}_{opt}^j),\\
                    \beta_{i,max}, i \in S^{k^\star},\\
                    {\displaystyle \frac{h_{si}}{h_{it}} \frac{1+\sum\limits_{j \in S^{k^\star}} \beta_{j,max}^2 h_{jt}^2}{\sum\limits_{j \in S^{k^\star}} h_{sj} \beta_{j,max} h_{jt}}, i \not\in S^{k^\star}},
                  \end{dcases}
\end{equation*}
where $S^{k^\star} = \{k^\star\} \cup \{i: \beta_{i}^{k^\star} \ge \beta_{i,max}, i \in \mathcal{N}\setminus\{k^\star\}\}$.
\end{pavikt}

Based on our approach in this section to compute the optimal ANC rate in the Gaussian diamond networks, in the next section we introduce a greedy scheme to bound from below the maximum end-to-end rate achievable with analog network coding in the general layered networks.

\section{General layered networks: a greedy scheme to lower bound the maximum ANC rate}
\label{sec:genNet}
\begin{figure}[!t]
\centering
\includegraphics[width=3.0in]{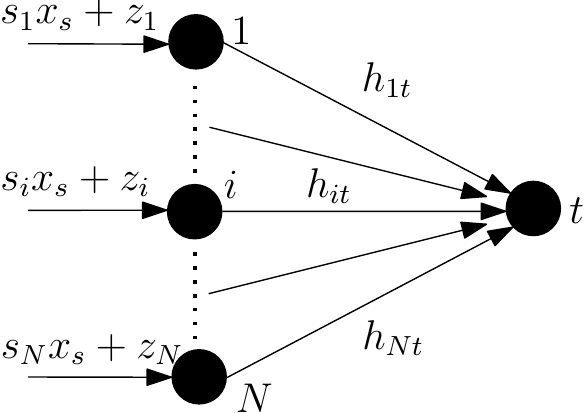}
\caption{A subnetwork of general layered network with $L$ relay layers, depicting $l^\textrm{th}$ layer with $N$ relay nodes and a node in the $l+1^\textrm{st}$ layer. The received signal component at node $i, 1 \le i \le N$, in the $l^\textrm{th}$ layer is denoted as $s_i x_s$, where $x_s$ is the source symbol and corresponding noise component is denoted as $z_i$.}
\label{fig:l2lplus1}
\end{figure}

In a general layered network with $L$ layers of relay nodes, consider layer $l, 1 \le l \le L$, and a node in the next $l+1^\textrm{st}$ layer, denoted as $t_{l+1}$ or with a little abuse of notation as $t$. This scenario is depicted in Figure~\ref{fig:l2lplus1}. For this subnetwork, for any scaling vector $\bm{\beta}$ we have
\begin{equation}
\label{eqn:subnetSNR}
SNR_t = \frac{P_s(\sum_{i=1}^N s_i \beta_i h_{it})^2}{\mathbb{E}(z_t+\sum_{i=1}^N z_i \beta_i h_{it})^2}
\end{equation}

Using \eqref{eqn:maxAFrate} the problem of computing the maximum ANC rate for this subnetwork can be formulated as
\begin{equation}
\label{eqn:l2lplus1prob}
\max_{0 \le \bm{\beta} \le \bm{\beta}_{max}} SNR_t,
\end{equation}
where $\bm{\beta} = (\beta_1, \ldots, \beta_N)$ and $\bm{\beta}_{max} = (\beta_{1,max} \ldots, \beta_{N,max})$ with $\beta_{i,max}^2 = P_i/\mathbb{E}(s_i x_s + z_i)^2, i \in {\mathcal N}, {\mathcal N} = \{1, \ldots, N\}$.

Equating the first-order partial derivatives of the objective function with respect to $\beta_i, i \in \mathcal{N}$, to zero, we get the following $N+1$ conditions for local extrema:
\begin{align}
&\sum_{i \in \mathcal{N}} s_{i} \beta_i h_{it} = 0 \label{eqn:gen_b1bN} \\
&\beta_i = \frac{s_i + s_i \mathbb{E}\big(\sum\limits_{j \in \mathcal{N} \setminus \{i\}} z_j \beta_{j} h_{jt}\big)^2 -  \alpha_i \gamma_i}{h_{it}(\alpha_i \mathbb{E}z_i^2 / \sigma^2 - s_i \gamma_i)} \label{eqn:gen_bi_ito_rest}
\end{align}
where
\begin{align*}
\alpha_i &= \sum\limits_{j \in \mathcal{N} \setminus \{i\}} s_j \beta_{j} h_{jt}, \quad\qquad \mbox{(signal component at destination $t$ from all the nodes except node $i$)}\\
\gamma_i &= \sum\limits_{j \in \mathcal{N} \setminus \{i\}} \beta_{j} h_{jt} \mathbb{E}(z_i z_j) / \sigma^2, \quad \mbox{(noise component at destination $t$ from all the nodes except node $i$)}
\end{align*}

As we established the extremal properties of conditions \eqref{eqn:b1bN} and \eqref{eqn:bi_ito_rest} in Section~\ref{sec:diamond}, we can also prove that condition \eqref{eqn:gen_b1bN} leads to the global minimum of the objective function in \eqref{eqn:l2lplus1prob} and the global maximum of the objective function occurs at one of the $N$ hyperplanes defined by $\beta_k = \beta_{k,max}$. Following a sequence of arguments similar to those used to establish Theorem~\ref{thrm:optBeta4diamond} for diamond networks, we can characterize the scaling vector for the nodes in the $l^\textrm{th}$ layer that optimally solve the problem~\eqref{eqn:l2lplus1prob} for the subnetwork under consideration. Note that in this subnetwork, the noises at different nodes in a relay layer are correlated, unlike the noises at relay nodes in the diamond network in Figure~\ref{fig:diamond}. This explains the difference between the $SNR$ expression in \eqref{eqn:subnetSNR} and the one in \eqref{eqn:diamondSNR} for the diamond network, and results in more complex analysis in the present case.

\begin{pavikl}
\label{lemma:optBeta4l2lplus1}
A scaling vector $\bm{\beta}_{opt} = (\beta_{1}^{opt}, \ldots, \beta_{N}^{opt})$ that maximizes the $SNR_t$ for any subnetwork, as in Figure~\ref{fig:l2lplus1}, of the general layered network with the relay nodes in $l^\textrm{th}$ layer performing analog network coding is given as
\begin{equation*}
\beta_{i}^{opt} = \begin{dcases}
                    \beta_{i,max}, i = k^\star, k^\star = \argmax_{\{\bm{\beta}^j:j \in \mathcal{N}\}} SNR_t(\bm{\beta}^j),\\
                    \beta_{i,max}, i \in S^{k^\star},\\
                    \frac{s_i + s_i \mathbb{E}\big(\sum\limits_{j \in S^{k^\star}} z_j \beta_{j,max} h_{jt}\big)^2 - \alpha_j \gamma_j}{h_{it}(\alpha_j\mathbb{E}z_i^2 / \sigma^2 - s_i \gamma_j)}, i \not\in S^{k^\star},
                  \end{dcases}
\end{equation*}
where $\bm{\beta}^j=(\beta_1^j, \ldots, \beta_N^j)$ with
\begin{align*}
\beta_i^j &= \begin{dcases}
               \beta_{j,max}, i = j\\
               \frac{s_i + \{s_i \mathbb{E}z_j^2 / \sigma^2 - s_j \mathbb{E}(z_i z_j) / \sigma^2\}\beta_{j,max}^2 h_{jt}^2}{h_{it}\{s_j \mathbb{E}z_i^2 / \sigma^2 - s_i \mathbb{E}(z_i z_j) / \sigma^2\}\beta_{j,max} h_{jt}}, i \neq j,
             \end{dcases}\\
\alpha_j &= \sum_{j \in S^{k^\star}} s_j \beta_{j,max} h_{jt}, \qquad\qquad \mbox{(signal component at destination $t$ from the nodes in $S^{k^\star}$)}\\
\gamma_j &= \sum_{j \in S^{k^\star}} \beta_{j,max} h_{jt} \mathbb{E}(z_i z_j) / \sigma^2, \quad \mbox{(noise component at destination $t$ from the nodes in $S^{k^\star}$)}
\end{align*}
and $S^{k^\star} = \{k^\star\} \cup \{i: \beta_{i}^{k^\star} \ge \beta_{i,max}, i \in \mathcal{N}\setminus \{k^\star\}\}$.
\end{pavikl}

Note that Lemma~\ref{lemma:optBeta4l2lplus1} reduces to Theorem~\ref{thrm:optBeta4diamond} when the noise components at the relay nodes are uncorrelated.

Using Lemma~\ref{lemma:optBeta4l2lplus1}, we can compute $\bm{\beta}_{l,opt}^{l+1, j}$, the scaling vector for the nodes in the $l^\textrm{th}$ layer that maximizes the received SNR for node $j, 1 \le j \le n_{l+1}$, in the $l+1^\textrm{st}$ layer. Among these $n_{l+1}$ scaling vectors for the nodes in the $l^\textrm{th}$ layer, let $\bm{\beta}_{l}^{low}$ denote the one that solves the following problem
\begin{equation}
\label{eqn:bestlbeta}
\bm{\beta}_{l}^{low} = \argmax_{\substack{\bm{\beta}_{l,opt}^{l+1, j} \\ 1 \le j \le n_l}} \prod_{k=1}^{n_{l+1}} (1+SNR_k)
\end{equation}
The following corollary of Lemma~2 in \cite{112agnihotriJaggiChen} establishes that among $n_{l+1}$ such scaling vectors, the scaling vector characterized by $\bm{\beta}_{l}^{low}$ computes the tightest lower bound for the optimal value of the objective function in \eqref{eqn:bestlbeta} as well as \eqref{eqn:eqProb}.

\begin{pavikc}[\hspace{-0.001cm}\cite{112agnihotriJaggiChen}, Lemma 2]
\label{cor:betterBeta}
Consider two scaling vectors $\bm{\beta}_l$ and $\bm{\hat\beta}_l$ for the nodes in $l^\textrm{th}$ layer. If $\prod_{k=1}^{n_{l+1}} (1+SNR_k)\big|_{\bm{\beta}_l} > \prod_{k=1}^{n_{l+1}} (1+SNR_k)\big|_{\bm{\hat\beta}_l}$, then $SNR_t(\bm{\beta}_l) > SNR_t(\bm{\hat\beta}_l)$.
\end{pavikc}

Computing $\bm{\beta}_{l}^{low}$ as above for each layer $l, 1 \le l \le L$, in conjunction with Corollary~\ref{cor:betterBeta}, allows us to construct a network-wide scaling vector $\bm{\beta}_{low} = (\bm{\beta}_{1}^{low}, \ldots, \bm{\beta}_{L}^{low})$ to compute a lower bound\footnote{Clearly, choosing $\bm{\beta}_{l}^{low}$ as in \eqref{eqn:bestlbeta} for each layer $l$ may lead, in general, to some performance loss at each layer as $\bm{\beta}_{l}^{low}$ may not be the optimal vector of the scaling factors for the nodes in the layer $l$ that solves
\begin{equation*}
\argmax_{0 \le \bm{\beta}_{l} \le \bm{\beta}_{l,max}} \prod_{k=1}^{n_{l+1}} (1+SNR_k)
\end{equation*}
The cumulative effect of this performance loss at each layer is that the end-to-end ANC rate computed at $\bm{\beta}_{low}$ may not lead to the optimal solution of problem~\eqref{eqn:maxAFrate}. However, our results in the next section show that for a large class of layered networks there is no loss in the optimality and for other layered networks, the loss is \textit{small} asymptotically in the network parameters.} to the optimal solution of \eqref{eqn:maxAFrate}. Formally, for a given layered network, $\bm{\beta}_{low}$ is constructed as follows.
\begin{pavikp}
\label{prop:lowBetaGenNet}
Consider a layered relay network of $L+2$ layers, with the source $s$ in layer `$0$', the destination $t$ in layer `$L+1$', and $L$ layers of relay nodes between them. The $l^\textrm{th}$ layer contains $n_l$ nodes, $n_0 = n_{L+1} = 1$. A network-wide scaling vector $\bm{\beta}_{low}=(\bm{\beta}_{1}^{low}, \ldots, \bm{\beta}_{L}^{low})$ that provides a lower bound to the optimal solution of \eqref{eqn:maxAFrate} for this network, can be computed recursively for $1 \le l \le L$ as
\begin{equation*}
\bm{\beta}_{l}^{low} = \argmax_{\substack{\bm{\beta}_{l,opt}^{l+1, j} \\ 1 \le j \le n_l}} \prod_{k=1}^{n_{l+1}}(1+SNR_{l+1, k}(\bm{\beta}_{1}^{low}, \ldots, \bm{\beta}_{l-1}^{low}, \bm{\beta}_{l,opt}^{l+1, j}))
\end{equation*}
Here $\bm{\beta}_{l}^{low} = (\beta_{l 1}^{low}, \ldots, \beta_{l n_l}^{low})$ is the vector of scaling factors for the nodes in the $l^\textrm{th}$ layer, and $\bm{\beta}_{l,opt}^{l+1, j}$ (computed using Lemma~\ref{lemma:optBeta4l2lplus1}) is the scaling vector for the nodes in the $l^\textrm{th}$ layer that maximizes the received SNR for node $j, 1 \le j \le n_{l+1}$, in layer $l+1$.
\end{pavikp}

In the next section we analyze the performance of the greedy scheme of the Proposition~\ref{prop:lowBetaGenNet} in the context of both a special class of layered networks and the general layered networks.

\section{Illustration}
\label{sec:exa}
We first demonstrate that the greedy scheme of Proposition~\ref{prop:lowBetaGenNet} allows us to exactly compute the optimal ANC rate for a broad class of layered networks. Then, we give an example to show that for the general layered networks, the proposed scheme leads to the optimal rates within a constant gap from the cut-set upper bound asymptotically in the source power.

\begin{figure}[!t]
\centering
\includegraphics[width=3.0in]{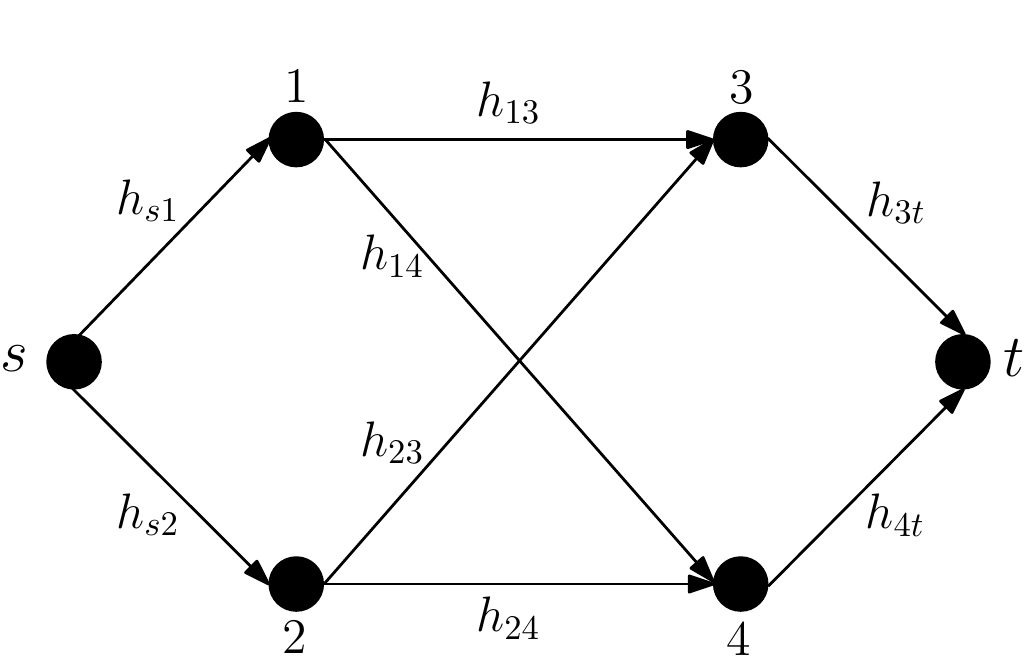}
\caption{General layered network with 2 relay layers between the source `s' and destination `t'. Each layer contains two relay nodes.}
\label{fig:layrdNetSpl_Exa}
\end{figure}

\textit{\textbf{Example 2} (A class of exactly solvable layered networks):} Let us consider a class of symmetric layered networks where the channel gains along all outgoing links from a node are equal. An instance of such a network is obtained from the network in Figure~\ref{fig:layrdNetSpl_Exa} when $h_{s1}=h_{s2}, h_{13}=h_{14}$, and $h_{23}=h_{24}$. An implication of this property of the channel gains is that the received SNRs at every node in a layer are equal: $SNR_{l, j}=SNR_l, 1 \le j \le n_l, 1 \le l \le L$. In this case, for each layer $l, 1 \le l \le L$, $\bm{\beta}_{l}^{low}$ computed in Proposition~\ref{prop:lowBetaGenNet} is equal to the optimal $\bm{\beta}_l^{opt}$ computed in \cite[Lemma 2]{112agnihotriJaggiChen}. Therefore, $\bm{\beta}_{low}$ is the optimal solution of problem \eqref{eqn:eqProb} for this class of networks.

Consider an instance of the network in Figure~\ref{fig:layrdNetSpl_Exa} when $h_{s1}=h_{s2}, h_{13}=h_{14}$, and $h_{23}=h_{24}$. Such an instance belongs to the class of symmetric networks we are concerned with in this example. Using Proposition~\ref{prop:lowBetaGenNet}, the optimal solution of problem \eqref{eqn:eqProb} for this instance is:
\begin{align*}
&\bm{\beta}_{opt} = \bigg(\beta_{1,max}, \frac{1+\beta_{1,max}^2 h_1^2}{h_2 \beta_{1,max} h_1}, \beta_{3,max}, \beta_{4,max}\bigg), \mbox{ where}\\
&\beta_{1,max}^2 = \frac{P_1}{h_0^2 P_s + \sigma^2} \\
&\beta_{3,max}^2 = \frac{P_3}{S^2 P_s + Z^2 \sigma^2}, \beta_{4,max}^2 = \frac{P_4}{S^2 P_s + Z^2 \sigma^2} \\
&S = h_0(\beta_{1,opt} h_1 + \beta_{2,opt} h_2), Z^2 = 1 + \beta_{1,opt}^2 h_1^2 + \beta_{2,opt}^2 h_2^2
\end{align*}
and we assume that $P_1 h_1^2 > P_2 h_2^2$. {\hspace*{\fill}~\IEEEQEDclosed\par}

\begin{figure}[!t]
\centering
\includegraphics[width=3.0in]{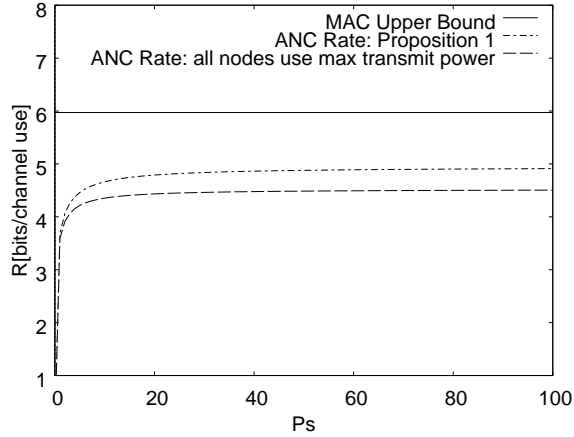}
\caption{Comparison of the ANC rate achievable with the scheme in Proposition~\ref{prop:lowBetaGenNet} with the MAC upper bound for the layered network in Figure~\ref{fig:layrdNetSpl_Exa} with $P_1=P_2=P_3=P_4=10$, $h_{14}=h_{24}=2$ and all other channel gains are equal to $10$. Also plotted is the ANC rate when the scaling factors for all relay nodes are set to their respective upper-bounds.}
\label{fig:rateComparison}
\end{figure}

{\textit{\textbf{Example 3} (General layered networks):}}
Let us consider the layered network of Figure~\ref{fig:layrdNetSpl_Exa}. We compute a lower bound to the optimal ANC rate for this network using the greedy scheme in the Proposition~\ref{prop:lowBetaGenNet} and compare it with the MAC upper bound in Figure~\ref{fig:rateComparison}. Also, plotted in this figure is the ANC rate achievable when the scaling factors for all relay nodes are set to their respective upper-bounds. We observe that in this case the ANC rate achieved with the greedy scheme of Proposition~\ref{prop:lowBetaGenNet} approaches the capacity within one bit when $P_s > 100$.

\section{Conclusion and Future Work}
\label{sec:conclFW}
We consider the problem of maximum rate achievable with analog network coding in general layered networks. Previously, this problem has been considered under certain assumptions on per node scaling factor and received SNR as without these assumptions the problem was presumed to be intractable. The key contribution of this work is a greedy scheme to exactly compute the optimal rates in a wider class of layered networks than those that can be addressed using prior approaches. In particular, using the proposed scheme for the Gaussian $N$-relay diamond network, to the best of our knowledge, we provide the first exact characterization of the optimal rate achievable with analog network coding. Further, for general layered networks, our scheme allows us to compute optimal rates at most a constant gap away from the cut-set upper bound asymptotically in the source power. In the future, we plan to extend this work to non-layered networks, and to construct the optimal distributed relay schemes.

\end{document}